\documentclass[fleqn,10pt]{wlscirep}
\usepackage[utf8]{inputenc}
\usepackage[T1]{fontenc}
\usepackage{bbm}
\usepackage[super]{nth}
\title{A New Data Integration Framework for Covid-19 Social Media Information}

\author[1]{Lauren Ansell}
\author[1,*]{Luciana Dalla Valle}
\affil[1]{University of Plymouth, School of Engineering,Computing and Mathematics, Plymouth, PL48AA, UK}
\affil[*]{luciana.dallavalle@plymouth.ac.uk}


\begin{abstract}
The Covid-19 pandemic presents a serious threat to people’s health, resulting in over 250 million confirmed cases and over 5 million deaths globally. 
To reduce the burden on national health care systems and to mitigate the effects of the outbreak, accurate
modelling and forecasting methods for short- and long-term health demand are needed to inform government
interventions aiming at curbing the pandemic.
Current research on Covid-19 is typically based on a single source of information, specifically on structured historical pandemic data.
Other studies are exclusively focused on unstructured online retrieved insights, such as data available from social media. 
However, the combined use of structured and unstructured information is still uncharted. 
This paper aims at filling this gap, by leveraging historical and social media information with a novel
data integration methodology. 
The proposed approach is based on vine copulas, which allow us to exploit the dependencies between different sources of information.
We apply the methodology to combine structured datasets retrieved from official sources and a big unstructured dataset of information collected from social media.
The results show that the combined use of official and online generated information contributes to yield a more accurate assessment of the evolution of the Covid-19 pandemic, compared to the sole use of official data.

\end{abstract}
\begin{document}

\flushbottom
\maketitle
% * <john.hammersley@gmail.com> 2015-02-09T12:07:31.197Z:
%
%  Click the title above to edit the author information and abstract
%
\thispagestyle{empty}

%\noindent Please note: Abbreviations should be introduced at the first mention in the main text – no abbreviations lists. Suggested structure of main text (not enforced) is provided below.

\section*{Introduction}

The outbreak of the Covid-19 disease infected and killed millions of people globally, resulting in a pandemic with enormous global impact. This disease affects the respiratory system, and the viral agent that causes it spreads through droplets of saliva, as well as through coughing and sneezing. As an extremely transmissible viral infection, Covid-19 is causing significant damage to countries' economies because of its direct impact on the health of citizens and the containment measures taken to curtail the virus. %\cite{pinheiro2021using}.
In the UK, Covid-19 had serious implications for people's health and the healthcare services, with more than 8 million confirmed cases and 150,000 deaths, as government figures show.
There is thus a widespread interest in accurately estimating and assessing the evolution of the pandemic over time.

Current studies on Covid-19 are typically based on a single source of information.
Most of them implemented quantitative analyses focusing on historical data to produce forecasts of the pandemic.
For example,  
clinical data of Covid‐19 patients was used to statistically analyse with meta‐analysis the clinical symptoms and laboratory results aiming at explaining the discharge and fatality rate \cite{li2020covid}.
Another study considered data on confirmed and recovered cases and deaths, the growth rate and the trend of the disease in Australia, Italy and the UK \cite{rahimi2021analysis}. The authors predicted epidemiology in the three countries with mathematical approaches based on susceptible, infected, and recovered (SIR) cases and susceptible, exposed, infected, quarantined, and recovered (SEIQR) cases, comparing them with the Prophet machine learning algorithm and the logistic regression.
Machine learning methods were also adopted 
to implement logistic regression and gradient boosted trees on health risk assessment questionnaires and medical claims diagnosis data to predict complications due to Covid-19 \cite{decaprio2020building}.
%\cite{ahmed2021framework} performed descriptive, diagnostic, predictive, and prescriptive analysis of the pandemic applying neural networks and other machine learning algorithms, focusing on different pandemic symptoms.
%\cite{pinheiro2021using} employed network analytics and machine learning models by using a combination of anonymized health and telecommunications data to understand the correlation between population movements and virus spread and to predict possible new outbreaks.

However, some authors in the literature criticize approaches which are exclusively based on official quantitative information.
Existing Covid-19 studies based on historical data and prediction models for the pandemic are ``poorly reported, at high risk of bias and underperforming'' \cite{wynants2020prediction, jewell2020predictive}. 

%Another strand of research focuses on different types of data, analysing textual information with Natural Language Processing (NLP) methods.
%For example, \cite{liu2021tracing} adopted the Latent Dirichlet Allocation (LDA) model, to allocate research articles into different research topics pertinent to Covid-19 according to their abstracts.
%A LDA model was also employed by \cite{wang2021information}, who applied it to the question and answer data about Covid-19 in Chinese online health communities.

Text data in the form of social media insights are used in the literature to evaluate and predict the progression of the Covid-19 pandemic.
For example, 
a lag correlation analysis was employed with data collected from Google Trends, Baidu Search Index and Sina Weibo Index \cite{li2020retrospective}.
Sina Weibo messages were also analysed in other studies \cite{liu2020characteristics, peng2020exploring, zhu2020limited}. 
A statistical analysis based on the Fisher exact test was carried out, 
the rates of death were calculated with the Kaplan-Meier method and risk factors for mortality were established using the multivariate Cox regression \cite{liu2020characteristics}. 
A kernel density analysis and an ordinary least square regression were implemented to identify the spatiotemporal distribution of Covid-19 cases in the main urban area of Wuhan, China \cite{peng2020exploring}. 
Descriptive statistics were calculated and a time series analysis was applied to the data \cite{zhu2020limited}.
Baidu Search Index information was analysed using different statistical methods, including the subset selection, forward selection, lasso regression, ridge regression and elastic net \cite{qin2020prediction}. 	
Google trends searches, Wikipedia page views and Twitter messages were gathered
implementing a regression analysis to show that online-retrieved information provided a leading indication of the number of people in the USA who became infected and die from the coronavirus \cite{o2020google}. 

However, contributions in the literature which are exclusively based on either official or online information as stand-alone sources are not taking into account the drawbacks affecting the data and the potential synergies between different data sources.
On the one hand, due to limited capacity of testing, official data on confirmed cases are unlikely
to reflect the true Covid-19 numbers. On the other hand, social media data are generated by
users on a voluntary basis and may not capture information about the entire population.
Therefore, predictive models built on a single source of information might generate biased results.

The goal of this paper is to develop a state-of-the-art data integration framework, leveraging the dependencies between historical and online data to provide more accurate evaluations 
%and predictions 
of the Covid-19 dynamics. The proposed approach exploits the synergies between official and online generated Covid-19-related data, yielding improved predictions of the pandemic compared to forecasts obtained solely with a single source of information.

Our approach is based on vine copulas, which are very flexible mathematical tools, able to correctly capture the dependence structures between different variables \cite{czado2019analyzing}.
Integration of different data sources using copulas and Bayesian networks was proposed in the literature \cite{dallavalle2014official, dalla2015official, dalla2018social}. However, the approach adopted by the authors was based on data calibration \cite{dalla2017dataint}.
In this paper, our aim is to propose a comprehensive novel data integration framework, able to improve data modelling and forecasting \cite{ansell2021social}.

So far, the application of copulas and vines to pandemic data has been limited to study the implications of Covid-19, especially in the financial field, rather than to directly calculate forecasts of pandemic trends.
For example, 
a Markov-switching dynamic copula with Student-t distribution was implemented to explore Covid-19 shock effects on energy markets \cite{maneejuk2021time}. 
A quantile regression model estimated via vine copula was used to show that speculation in energy and precious metal futures are more prevalent in crisis periods such as the Covid-19 pandemic \cite{sifat2021covid}.

The vine copula approach allows us to obtain the joint multivariate distribution of the marginals, embedding the actual dependence structure between them.
The marginals can belong to any distributional family and can potentially be all different.
Vine copulas use bivariate (pair) copulas as building blocks. Since there exist a wide variety of bivariate copula families, able to capture as symmetric as well as asymmetric dependencies, the vine copula approach is highly flexible, allowing us to model virtually any type of dependence structures between each pair of variables\cite{czado2019analyzing}.

This paper proposes a novel data integration methodology based on vine copulas, able to exploit the synergies between two different types of information: (a) official and (b) social media data. 
As official information (a) we consider structured historical pandemic data and, as social media data (b), we consider a large unstructured online-retrieved dataset on Covid-19, relevant to the UK geographical area.
We compare three models: (i) the flexible vine copula approach, where all pair copulas are different and are able to capture any type of dependencies between variables; (ii) the Gaussian copula approach, where all pair copulas are set to be equal to the Gaussian copula, and hence are only able to capture symmetric dependence; (iii) the independence approach, where we assume absence of dependence between the variables (in particular, no dependence between the official and social media data). 
Models (ii) and (iii) are traditional approaches, which impose restrictions on the relationships between variables.
In particular, the Gaussian model (ii) assumes the existence of relations between official (a) and social media data (b), yet it allows only the traditional Gaussian family of distributions to describe this relationship. 
The independence model (iii), instead, does not allow the existence of any form of dependence between the variables and, in particular, between those derived by official (a) and social media (b).
We are particularly interested in assessing the contribution of online-generated data (b) to the evaluation and prediction of the Covid-19 pandemic, hence in the performance of models (i) and (ii) compared to model (iii).
Vine copulas allow us to compute predictions based on the joint multivariate distribution of the marginals. Differently from regression models, all marginals in vine copula models hold the same role and can be used in turn to predict selected variables. Here, we are specifically interested to evaluate how social media information (b) perform in predicting official pandemic data (a).
%This paper will be the first to propose statistical methodology based on vine copulas, able to exploit and integrate official and social media data, to accurately model the spread of Covid-19.
%The methodology will be applied to structured historical pandemic data and to a large unstructured online-retrieved dataset, relevant to the UK geographical area.
Our results will show that the vine copula  approaches (i) and (ii) perform  better  than the independence traditional  approach (iii), which does not take into account associations between official and on-line information to estimate and predict the Covid-19 dynamic.

%The  remainder  of  the  paper  is  organised as  follows. The second section %\ref{sec:data} 
%describes the different data sources used in the analysis;
%The third section %\ref{sec:method} 
%illustrates
%the vine copula methodology; the fourth section %\ref{sec:analysis} 
%reports the results of the analysis; finally,
%concluding remarks are presented in the last section. %\ref{sec:Conclusions}.

%%%%%%%%%%%%%%%%%%%%%%%%%%%%%%%%%%%
%%     Datasets 			%%%
%%%%%%%%%%%%%%%%%%%%%%%%%%%%%%%%%%%

\section*{Dataset}\label{sec:data}

The structured and unstructured data used in this paper were collected daily between the 21st April 2020 and the 9th May 2021.
As structured official data (a), we considered the following 6 variables: the number of new admitted patients (\texttt{Admissions}), the number of hospital cases (\texttt{Hospital}), the number of patients on ventilation (\texttt{ICU\_Beds}), the number of tests (\texttt{VirusTests}), the number of positive cases (\texttt{Cases}) and the number of deaths (\texttt{Deaths}).
The first four data variables were gathered from the UK Government dashboard, % \cite{gov.uk}, %(Available at the website \url{https://coronavirus.data.gov.uk/}), 
while the last two variables were downloaded from the Johns Hopkins University database. % \cite{jhu.edu}. %(Available at the website \url{https://coronavirus.jhu.edu/}).
This information was available in cumulative form, therefore to obtain the daily time series, the previous days total was subtracted from the current days total.
As unstructured online data (b), we collected Google Trends information on the number of searches for the keywords \textit{Covid-19}, \textit{coronavirus}, \textit{first wave}, \textit{second wave} and \textit{variant}, using the \texttt{gtrendsR} package from the R software \cite{gtrendsR, Rsoftware}.
In particular, the function \texttt{gtrends} from the R package allows us to specify the Google Trends query keywords, the geographical region (in this case ``GB'') and the timespan of the query. 
In addition, we retrieved Twitter messages containing the same keywords used to perform Google Trends searches, using the \texttt{rtweet} R package \cite{rtweet-package}.
More specifically, we used the \texttt{search\_tweets} function of the R package, which allows us to specify the keywords used to filter and select tweets.
Three batches of 18,000 tweets were collected 3 times a day, everyday, due to
the restrictions on the maximum number of tweets to be downloaded in each query by \texttt{rtweet}.
Tweets can also be collected directly via Twitter’s standard
search application programming interface (API), with the advantage that the user is not bound to any specific software, as shown by other researchers in the literature who also employ different Covid-19–related search words to filter relevant tweets\cite{lwin2022evolution}.
For our analysis, we proceeded with a meticulous data cleansing, which lead to a considerable reduction of the final number of tweets that we included in our dataset.
We carefully discarded any tweets not directly related to the Covid-19 pandemic, 
since, for example, the keywords \textit{first wave}, \textit{second wave} and \textit{variant} led to some irrelevant tweets and noisy information. 
We removed tweets written from outside the UK and those produced from locations with less than 10 tweets. 
We also deleted any duplicate tweets and those sent by automated accounts which contained factual information about daily case numbers or retweets of news stories.
Finally, we removed tweets directly addressed to foreign political leaders or politicians, obtaining a final large Twitter dataset of 577,231 tweets.
From the Twitter data, we considered the total number of tweets as well as the sentiment scores calculated using two different lexicons: Bing and Afinn \cite{hu2004mining}, which are available in the R \texttt{tidytext} package \cite{tidytext}.
The Bing lexicon splits words into positive or negative. The Bing sentiment score for each tweet is calculated by counting the number of positive words used in each tweet and subtracting from this the number of negative words.
The Afinn lexicon scores words between $\pm 5$. The Afinn sentiment score is calculated by multiplying the score of each word by the number of times it appears in the tweet; these scores are then summed to derive the overall sentiment score.
Therefore, from online-gathered information (b) we obtained the following 4 variables: Afinn sentiment score (\texttt{Afinn}), Bing sentiment score (\texttt{Bing}), Google trends (\texttt{Google}) and the total number of tweets (\texttt{Tweets}).

%\begin{figure}[h]
%    \centering
%    \includegraphics[width=1\textwidth]{Trace_plot.png} 
%    \caption{Trace plots of Covid-19 data.}
 %   \label{fig:trace_plot}
%\end{figure}

%Figure \ref{fig:trace_plot} shows the trace plots of the Covid-19 official and social media times series.
%The panels depict the variables (from top to bottom) \texttt{Admissions}, the Afinn sentiment scores (\texttt{Afinn}), the Bing sentiment scores (\texttt{Bing}), \texttt{Cases}, \texttt{Deaths}, the Google Trends searches (\texttt{Google}), \texttt{Hospital}, \texttt{ICU\_Beds}, the total number of Tweets (\texttt{Tweets}) and \texttt{VirusTests}.
%We notice a higher volume of online messages produced at the beginning of the collection period and spikes corresponding to periods of more heated online discussions. 
%In addition, the daily reporting of official structured data shows high variability and there is often a lag in reporting in the UK government figures due to figures being under reported at the weekend. The highlighted drawbacks of the official data could be overcome by data integration with social media information, which do not suffer from reporting lags.

%%%%%%%%%%%%%%%%%%%%%%%%%%%%%%%%%%%
%%     Methodology 			%%%
%%%%%%%%%%%%%%%%%%%%%%%%%%%%%%%%%%%

\section*{Methodology}\label{sec:method}

The copula is a function that allows us to bind together a set of marginals, to model
their dependence structure and to obtain the joint multivariate distribution \cite{joe1997multivariate, nelsen2007introduction, dalla2017copulas, dalla2017copulasfin}.
Sklar’s theorem \cite{sklar1959fonctions} is the most important result in copula theory. It states that, given a vector of random variables $\textbf{X}=(X_1, \ldots, X_d)$, with $d$-dimensional joint cumulative distribution function $F(x_1, \ldots,x_d)$ and marginal cumulative distributions (cdf) $F_j(x_j)$, with $j=1, \ldots, d$, a $d$-dimensional copula $C$ exists, such that
$$
F(x_1, \ldots,x_d) = C(F_1(x_1), \ldots, F_d(x_d); \boldsymbol{\theta}),
$$
where $F_j(x_j) = u_j$, with $u_j \in [0,1]$ are called \textit{u-data}, and $\boldsymbol{\theta}$ denotes the set of parameters of the copula.
The joint density function can be derived as
$$
f(x_1, \ldots,x_d) = c(F_1(x_1), \ldots, F_d(x_d); \boldsymbol{\theta}) \cdot f_1(x_1) \cdot \cdot \cdot f_d(x_d),
$$
where $c$ denotes the $d$-variate copula density.
The copula allows us to determine the joint multivariate distribution and to describe the dependencies among the marginals, that can potentially be all different and can be modelled using distinct distributions.
As it is clear from the previous equations, differently from regression models, the variables $X_1, \ldots, X_d$ hold the same role and all of them, in turn, can be used to calculate predictions.

In this paper, we adopt the 2-steps inference function for margins (IFM) approach \cite{joe1996estimation}, estimating the marginals in the first step, and then the copula, given the marginals, in the second step.

%%%%%%%%%%%%%%%%%%%%%%%%%%%%%%%%%%%
%%     Marginal Models 			%%%
%%%%%%%%%%%%%%%%%%%%%%%%%%%%%%%%%%%

\subsection*{Marginal Models}\label{sub:marginals}

Given the different characteristics of the ten marginals, we fitted different models for each of the ten time series.
Further, we extracted the residuals $\varepsilon_j$, with $j = 1, \ldots, d$, from each marginal model and we applied the relevant distribution functions to get the \textit{u-data} $F_j(\varepsilon_j) = u_j$ to be plugged into the copula.
%According to the characteristics of each of the six time series marginals for each location, we used different models which provided the best fit to the data.
%Then, the \textit{u-data} to be plugged into the copula were obtained first extracting the residuals $\varepsilon_j$, with $j = 1, \ldots, d$, from each marginal model and then applying the relevant distribution function $F_j(\varepsilon_j) = u_j$.

\begin{itemize}
\item \textbf{New admitted patients (\texttt{Admissions})}

The best fitting model for the \texttt{Admissions} marginal was the SHASHo2 model. 
This model belongs to the family of GAMLSS distributions, which stands for Generalised Additive Models for Location, Scale and Shape. GAMLSS are very flexible models, which include a wide range of continuous and discrete distributions. %\cite{stasinopoulos2017flexible}.
The SHASHo2 model is also known as Sinh-Arcsinh original type 2 distribution and depends on four parameters: $\mu$ the location parameter, $\sigma$ the scaling parameter, $\nu$ the skewness parameter and $\tau$ the kurtosis parameter \cite{jones2009sinh}.
%The probability density function (pdf) of the SHASHo2 model is given by
%\begin{equation}\label{eq:SHASHo2}
%f_X(x|\mu,\sigma,\nu,\tau)= \frac{c}{\sqrt{2 \pi}}\frac{1}{\sqrt{1+z^2}} \exp\left(-\frac{r^2}{2}\right)
%\end{equation}
%for $ -\infty < x < \infty$, $ -\infty < \mu < \infty$, $\sigma>0$, $ -\infty < \nu < \infty$ and $\tau>0$, where $z=(x-\mu)/(\sigma \tau^2)$, $r = \sinh [ \tau \sinh^{-1} (z)-\nu ]$ and $c = \cosh [ \tau \sinh^{-1} (z)-\nu ]$.
We assumed that the parameter $\mu$ of the SHASHo2 model is related to time, as explanatory variable, through an appropriate link function, with coefficient $\beta$ \cite{rigby2005generalized}.
%, rigby2019distributions}).

\newpage

\item \textbf{Afinn sentiment score (\texttt{Afinn})}

We fitted the \texttt{Afinn} marginal with a reparametrized version of Skew Student $t$ type 3 model (SST), which, similarly to the previous marginal, belongs to the family of GAMLSS distributions and depends on four parameters: the mode ($\mu$), scaling ($\sigma$), skewness ($\nu$) and kurtosis ($\tau$) \cite{fernandez1998bayesian}. 
%The pdf for the SST model is 
%\begin{equation*}
% f_X(x|\mu, \sigma, \nu, \tau)= \frac{c}{\sigma} \left[1+\frac{z^2}{\tau} \left( \nu^2 \,\, \mathbbm{1}(x<\mu)+\frac{1}{\nu^2} \,\, \mathbbm{1}(x \geq\nu) \right) \right]^{-\frac{\tau+1}{2}}
%\end{equation*}
%for $-\infty< x < \infty$, $-\infty<\mu <\infty$, $\sigma >0$, $\nu>0$ and $\tau>0$, where $z=(x-\mu)/\sigma$, $c=2\nu[(1+\nu^2)B(1/2,\tau/2)\tau^{1/2}]^{-1}$, $B$ is the beta function and $\mathbbm{1}(\cdot)$ is the indicator function. 
Similarly to the SHASHo2 model, for the SST model we assumed that the parameter $\mu$ is related to time, as explanatory variable, through an appropriate link function, with coefficient $\beta$.

\item \textbf{Bing sentiment score (\texttt{Bing})}

The best model for \texttt{Bing} was the Normal-Exponential-$t$ (NET) distribution.
This is again a four parameter continuous distribution belonging the GAMLSS family \cite{rigby1994robust}. The parameters are: mean ($\mu$), scaling ($\sigma$), first kurtosis parameter ($\nu$) and second kurtosis parameter ($\tau$). 
%The pdf of the NET model is given by
%\begin{equation*}
%f_X(x|\mu, \sigma, \nu, \tau)= \frac{c}{\sigma}
%\begin{cases} 
%\exp \left(-\frac{z^2}{2} \right) & \text{when } |z| \leq \nu \\
%\exp \left( -\nu |z| + \frac{\nu^2}{2} \right)         & \text{when } \nu < |z| \leq \tau\\
%\exp \left( -\nu\tau \log \left( \frac{|z|}{\tau} \right) -\nu\tau+\frac{\nu^2}{2} \right) & \text{when  } |z|>\tau
%\end{cases}
%\end{equation*}
%for $-\infty < x < \infty$, $-\infty <\mu < \infty$, $\sigma >0$, $\nu>0$, $\tau> \max(\nu, \nu^{-1})$, where $z=(x-\mu)/\sigma$ and $c=(c_1+c_2+c_3)^{-1}$, $c_1=\sqrt{2\pi}[2\Phi(\nu)-1]$, $c_2=\frac{2}{\nu}\exp \left(-\frac{\nu^2}{2}\right)$ and $c_3=\frac{2}{(\nu\tau-1)\nu}\exp \left(-\nu\tau+\frac{\nu^2}{2}\right)$, with $\Phi$ the cumulative distribution function of the standard normal distribution.
As with the previous marginals, we assumed that the parameter $\mu$ of the NET model is related to time.

\item \textbf{Number of positive cases (\texttt{Cases})}

We fitted the \texttt{Cases} marginal with an ARIMA-GARCH model with Student's t innovations.
This model combines the features of the autoregressive integrated moving average (ARIMA) model with the generalized autoregressive conditional heteroskedastic (GARCH) model, allowing us to capture time series volatility over time \cite{hyndman2018forecasting}. The GARCH model is typically denoted as GARCH($p$, $q$), with parameters $p$ and $q$, where $p$ is the number of lag residuals errors and $q$ is the number of lag variances.
%The ARIMA($p$, $d$, $q$)-GARCH($p$, $q$) model can be expressed as:
%$$
%    y_t = a + \sum_{i=1}^p \phi_i y_{t-i} + \sum_{i=1}^q \theta_i \varepsilon_{t-i} + \varepsilon_t 
%$$
%\begin{equation}\label{eq:ARIMA-GARCH}
%    \varepsilon_t = \sqrt{\sigma_t} z_t \hspace{1cm} \sigma^2 = \omega + \sum_{i=1}^p \alpha_i \varepsilon_{t-i}^2 + \sum_{i=1}^q \beta_i \sigma_{t-i}^2 
%\end{equation}
%where the first line is the ARIMA part of the model, while the second line is the GARCH part of the model.
%Also, $y_t =(1-B)^d x_t$, $x_t$ are the original data values, $B$ is the backshift operator,
%$a$ is a constant, $\phi_i$ are the autoregressive parameters, $\theta_i$ are the moving average parameters;
%$\alpha_i$ and $\beta_i$ are the parameters of the GARCH part of the model, and $\varepsilon_t$ follows a Student's t distribution.

\item \textbf{Number of deaths (\texttt{Deaths})}

The best model for the \texttt{Deaths} marginal was the SHASHo model, whose acronym stands for Original Sinh-Arcsinh distribution. This model is very similar to the SHASHo2. %, represented in Eq.\eqref{eq:SHASHo2}.
%The pdf of the SHASHo model is 
%\begin{equation}\label{eq:SHASHo}
%f_X(x|\mu,\sigma,\nu,\tau)= \frac{c}{\sigma} \frac{\tau}{\sqrt{2 \pi}}\frac{1}{\sqrt{1+z^2}} \exp\left(-\frac{r^2}{2}\right)
%\end{equation}
%for $ -\infty < x < \infty$, $ -\infty < \mu < \infty$, $\sigma>0$, $ -\infty < \nu < \infty$ and $\tau>0$, where $z=(x-\mu)/(\sigma \tau)$, $r = \sinh [ \tau \sinh^{-1} (z)-\nu ]$ and $c = \cosh [ \tau \sinh^{-1} (z)-\nu ]$.
As for the other marginals fitted with GAMLSS-type models, we assumed that the parameter $\mu$ of the SHASHo depends on time.

\item \textbf{Google trends (\texttt{Google})}

Since \texttt{Google} includes values equal to zero, we fitted a Tweedie Generalised Linear Model for this marginal \cite{dunn2018generalized}. 
The Tweedie distribution has nonnegative support and can have a discrete mass at zero, making it useful to model responses that are a mixture of zeros and positive values.
%The Tweedie distribution belongs to the exponential family, assuming $E[x_t] = \mu_t^q = \zeta_t b$ as mean and $Var[x_i]=\phi \mu_t^p$ as variance, for $t = 1,\ldots, T$,
%\begin{equation*}
%    E[x_t] = \mu_t^q = \zeta_t b, \hspace{1cm} Var[x_i]=\phi \mu_t^p
%\end{equation*}
%where $\zeta_t$ denotes the time covariate, $b$ is the associated regression coefficient, $\phi$ is the dispersion parameter and $p$ and $q$ are extra parameters that control the mean and variance of the distribution, respectively.

\item \textbf{Number of hospital cases (\texttt{Hospital})}

The best model for the \texttt{Hospital} marginal was the ARIMA-GARCH model with Student's t innovations. 
%, as illustrated in Eq.\eqref{eq:ARIMA-GARCH}.

\item \textbf{Number of patients on ventilation (\texttt{ICU\_Beds})}

The best fitting model for the \texttt{ICU\_Beds} marginal was the SHASHo model.
%, with pdf given by Eq.\eqref{eq:SHASHo}.

\item \textbf{Total number of tweets (\texttt{Tweets})}

We fitted the \texttt{Tweets} marginal with a Skew Exponential Power type 4 (SEP4) model, which is a four parameter distribution belonging to the GAMLSS family \cite{rigby2005generalized}. 
%This is a ``spliced-shaped'' distribution with the following pdf
%\begin{equation*}
% f_X(x|\mu, \sigma, \nu, \tau)= \frac{c}{\sigma} \left[ \exp(-|z|^\nu) \,\, \mathbbm{1} (x < \mu)+\exp(-|z|^\tau) \,\, \mathbbm{1} (x \geq \mu) \right]  
%\end{equation*}
%for $-\infty< x < \infty$, $-\infty<\mu <\infty$, $\sigma >0$, $\nu>0$ and $\tau>0$, where $z=(x-\mu)/\sigma$, $c=\left(\Gamma (1+\nu^{-1})+\Gamma (1+\tau^{-1})\right)^{-1}$, $\Gamma$ is the gamma function and $\mathbbm{1}(\cdot)$ is the indicator function. Note that $\mu$ is the mode of $Y$.
Here we assumed that the parameter $\mu$ is related to time, as explanatory variable.

\item \textbf{Number of tests (\texttt{VirusTests})}

The best model for the \texttt{VirusTests} marginal was the  ARIMA-GARCH model with Student’s t innovations,
%, as illustrated in Eq.\eqref{eq:ARIMA-GARCH}, 
fitted on the the number of test adjusted by $1000$.

\end{itemize}

%%%%%%%%%%%%%%%%%%%%%%%%%%%%%%%%%%%
%%     Vine Copula Model 		%%%
%%%%%%%%%%%%%%%%%%%%%%%%%%%%%%%%%%%

\subsection*{Vine Copula Model}\label{sub:vinecopula}

A \textit{vine copula} (or \textit{vine}) represents the pattern of dependence of multivariate data via a cascade of bivariate copulas, allowing us to construct flexible high-dimensional copulas using only bivariate copulas as building blocks. 
The vine structure is highly flexible, since the bivariate copula densities can belong to any family %(e.g. Gaussian, Student’s t, Clayton, Gumbel, Frank, Joe, BB1, BB6, BB7, BB8, etc.).
\cite{czado2019analyzing}.

Two particular types of vines are the Gaussian vine (that we named model (ii)) and the Independence vine (model (iii)). The first one is constructed using solely Gaussian bivariate pair-copulas as building blocks, such that each conditional bivariate copula density %$c_{X_J,\nu_{\ell};\textbf{V}_{-\ell}}(\cdot, \cdot)$ described in Eq.\eqref{eq:condgen} 
is a Gaussian copula. The second type is the independence vine, which is constructed using only independence pair-copulas, that are simply given by the product of the marginal distributions of the random variables. In this latter case each conditional bivariate copula density %$c_{X_J,\nu_{\ell};\textbf{V}_{-\ell}}(\cdot, \cdot)$ described in Eq.\eqref{eq:condgen} 
is an Independence copula, implying absence of dependence between the variables.

%Pair copula constructions can be represented through a graphical model called \textit{regular vine} (R-vine).
%An R-vine $\mathcal{V}(d)$ on $d$ variables is a nested set of trees (connected acyclic graphs) $T_1, \ldots, T_{d-1}$, where the variables are represented by nodes linked by edges, each associated with a certain bivariate copula in the corresponding pair copula construction. The edges of tree $T_k$ are the nodes of tree $T_{k+1}$, $k = 1, \ldots, d-1$. 
%Two edges can share a node in tree $T_k$ without the associated nodes in tree $T_{k+1}$ being connected. In an R vine, two edges in $T_k$ which become two nodes in tree $T_{k+1}$, can only share an edge if in tree $T_k$ the edges shared a common node, but they are not necessarily connected by an edge. 
%\begin{figure}[h]
%    \centering
 %   \includegraphics[width=0.9\textwidth]{6dimVine.png}
 %   \caption{Six-dimensional R-vine graphical representation. \textit{Source: \cite{czado2019analyzing}}}
%    \label{fig:6dimVine}
%\end{figure}
%Figure \ref{fig:6dimVine} shows the 6-dimensional R-vine represented in Eq.\eqref{eq:pcc}. Each edge corresponds to a pair copula density (possibly belonging to different families) and the edge label corresponds to the subscript of the pair copula density, e.g. edge $2,4;1,3$ corresponds to the copula $c_{2,4;1,3}$.

In order to estimate the vine, its structure as well as the copula parameters have to be specified. 
A sequential
approach is generally adopted to select a suitable vine decomposition, specifying the first tree and then proceeding similarly for the
following trees. For selecting the structure of each tree, we followed the approach
based on the maximal spanning tree
algorithm \cite{aas2009pair, dissmann2013selecting}.
%This algorithm defines a tree on all nodes (named spanning tree), which
%maximizes the sum of absolute pairwise dependencies, measured, for example, by
%Kendall's $\tau$. This specification allows us to capture the strongest dependencies in the first tree and to obtain a more parsimonious model.
Given the selected tree structure, a copula family for each pair of variables is
identified using the Akaike Information Criterion (AIC), or the Bayesian Information
Criterion (BIC). This choice is typically made amongst a large set of families, comprising elliptical
copulas (Gaussian and Student's t) as well as Archimedean copulas (Clayton, Gumbel, Frank and
Joe), their mixtures (BB1, BB6, BB7 and BB8) and their rotated versions, to cover a large range of possible
dependence structures \cite{joe1997multivariate, nelsen2007introduction}.
The copula parameters $\boldsymbol{\theta}$ for each pair-copula in the vine 
are estimated using the maximum likelihood (MLE) method \cite{aas2009pair}.
The vine estimation procedure is repeated for all the trees, until the vine is completely
specified.

\subsection*{Out-of-sample predictions}\label{sec:out-of-sample}

In order to evaluate the suitability of the proposed vine copula model in relation to other methods, we produced one-day-ahead out-of-sample predictions and we compared them to the original data. Let $\textbf{X} = \{ \textbf{X}_t; t = 1, .., T \}$ be the $10$-dimensional time series of Covid-19 and social media data. Our aim is to forecast $\textbf{X}_{T+1}$ based on the information available at time $T$ \cite{simard2015forecasting}.
Before fitting the vine, we extracted the residuals from the marginals and obtained the \textit{u-data}. Next, after fitting the vine, we simulated $M$ realizations from the vine copula. Hence, we
calculated the predicted values for each simulation, using the inverse cdf and the relevant fitted marginal models.
More precisely, we applied the inverse transformation to the $M$ realizations from the vine copula to obtain the residuals which we then plugged into the marginal models to get the predicted values of the official variables. 
Then, we calculated the average prediction for all simulations $\hat{\textbf{X}}_{T+1}^{Avg}$ and use it to forecast $\textbf{X}_{T+1}$.
The prediction interval of level $(1 - \alpha) \in (0, 1)$ for $\textbf{X}_{T+1}$  was calculated by taking the estimated
quantiles of order $\alpha/2$ and $1 - \alpha/2$ amongst the simulated data. We denote by $\hat{\textbf{X}}_{T+1}^l$
and $\hat{\textbf{X}}_{T+1}^u$
the lower and upper values of the prediction intervals.

In order to compare and contrast the accuracy of predictions for different models, we made use of two indicators: the mean squared error (MSE) to evaluate point forecasts and the mean interval score (MIS) \cite{gneiting2007strictly}, to assess the accuracy of the prediction intervals.
The MSE for each variable $j = 1,\ldots, d$ was calculated as follows
$$
\mbox{MSE}_j = \frac{1}{S} \sum_{t=T+1}^{T+S} 
(x_{t,j} - \hat{x}_{t,j})^2
$$
where $x_{t,j}$ is the observed value for each variable at each time point $t$, $\hat{x}_{t,j}$ is the corresponding predicted value, $T+1$ denotes the first predicted date, while $T+S$ indicates the last predicted date. 
The 95\% MIS for each variable, at level $\alpha=0.05$, was computed as
$$
\mbox{MIS}_j = \frac{1}{S} \sum_{t=T+1}^{T+S} \left[ (\hat{x}_{t,j}^u - \hat{x}_{t,j}^l) + \frac{2}{\alpha} (\hat{x}_{t,j}^l - x_{t,j}) \mathbbm{1}(x_{t,j} < \hat{x}_{t,j}^l) + \frac{2}{\alpha} (x_{t,j} - \hat{x}_{t,j}^u) \mathbbm{1}(x_{t,j} > \hat{x}_{t,j}^u) \right]
$$
where $\hat{x}_{t,j}^l$ and $\hat{x}_{t,j}^u$ denote, respectively, the lower and upper limits of the prediction intervals for each variable at each time point, and $\mathbbm{1}(\cdot)$ is the indicator function.

\subsection*{Code Availability}

The code that implements the methodology described in the paper is available in
the GitHub repository \url{https://github.com/laurenansell/A-New-Data-Integration-Framework-for-Covid-19-Social-Media-Information}

%%%%%%%%%%%%%%%%%%%%%%%%%%%%%%%%%%%
%%     Result Analysis 			%%%
%%%%%%%%%%%%%%%%%%%%%%%%%%%%%%%%%%%

\section*{Result Analysis and Discussion}\label{sec:analysis}

We now present the results of the analysis of the official and online-retrieved Covid-19 data.

\subsection*{Twitter Wordclouds}

First, as an exploratory tool, we analysed the information gathered on Twitter, cleaning and stemming the tweets and producing graphical representations of the data using wordclouds. 
We point out that the analysis of wordclouds does not contribute to prediction, but it is a useful step to provide insights on the online-retrieved text information.

\begin{figure}[ht]
    \centering
    \includegraphics[width=\linewidth]{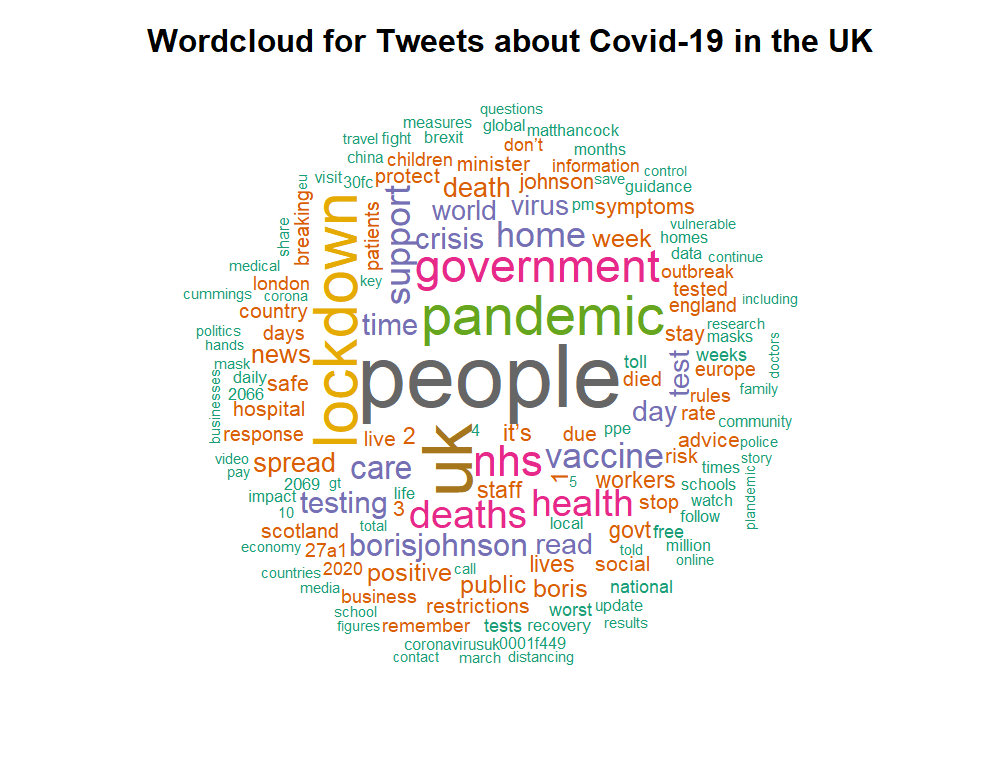} 
    \caption{Wordcloud of the UK Covid-19 data. \textit{Figure created by the authors using the package \texttt{wordcloud} of the R software version 4.2.2 \url{https://cran.r-project.org/}}}
    \label{fig:wordcloud}
\end{figure}

Figure \ref{fig:wordcloud} displays the wordcloud obtained from the collected tweets discussing Covid-19 in the UK. 
The most frequent words are related to ``people'' and the effects of the pandemic on them. We can also notice the names of the most prominent politicians and words related to political decisions.

\begin{figure}[ht]
    \centering
    \includegraphics[width=\linewidth]{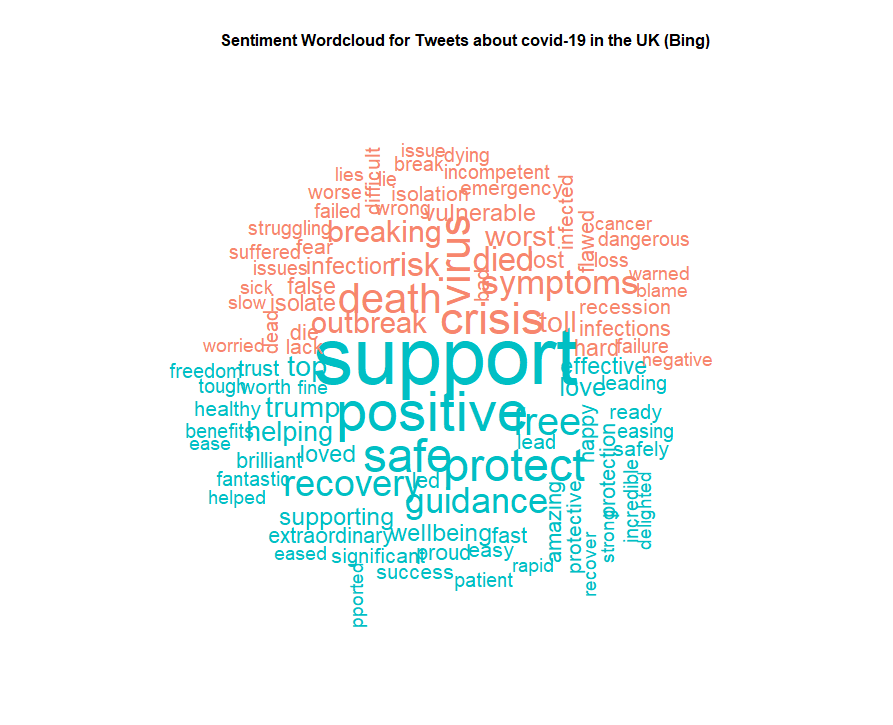} 
    \caption{Sentiment wordcloud for the UK Covid-19 data obtained with the Bing method. Positive words are in blue, negative words are in pink. \textit{Figure created by the authors using the package \texttt{wordcloud} of the R software version 4.2.2 \url{https://cran.r-project.org/}}}
    \label{fig:bingwordcloud}
\end{figure}

Figure \ref{fig:bingwordcloud} shows the sentiment wordcloud created from the collected tweets, obtained with the Bing method. This data visualization highlights the positive words in blue and the negative words in pink.
The most popular positive words are related to the ``support'' received troughout the pandemic, while the most popular negative words are related to the worst consequences of Covid-19 on the health of individuals.

%\begin{figure}[H]
%    \centering
%    \includegraphics[width=0.85\textwidth]{Bing_bar.png} 
%    \caption{Distribution of Bing sentiment for the UK Covid-19 data.}
%    \label{fig:bingbarchart}
%\end{figure}

%\begin{figure}[H]
%    \centering
%    \includegraphics[width=0.85\textwidth]{Afinn_bar.png} 
%    \caption{Distribution of Afinn sentiment for the UK Covid-19 data.}
%    \label{fig:afinnbarchart}
%\end{figure}

\subsection*{Marginals Estimation}

\begin{figure}[ht]
    \centering
    \includegraphics[width=0.6\textwidth]{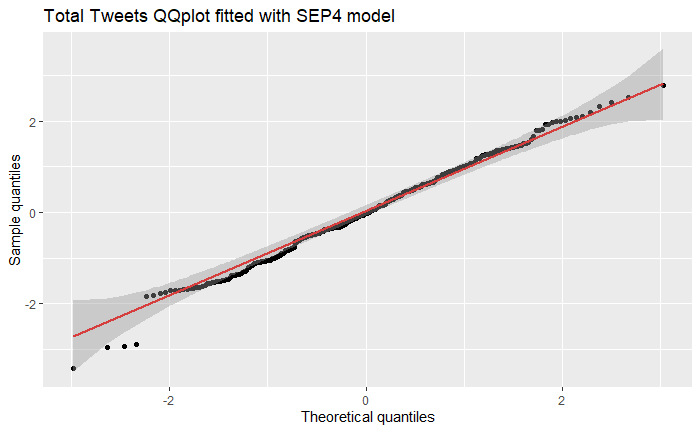} \\
    \includegraphics[width=0.6\textwidth]{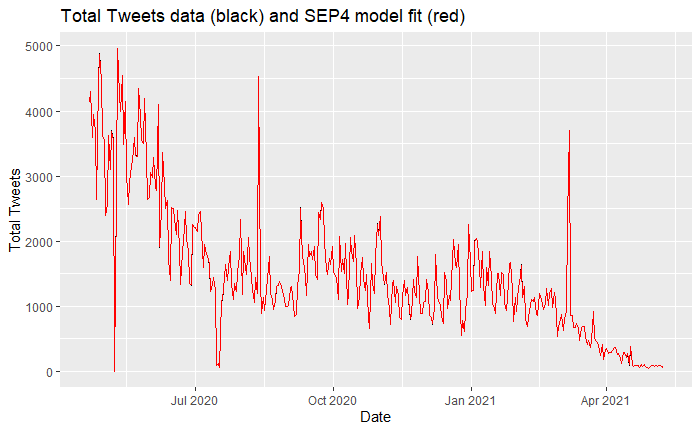}\\
    \includegraphics[width=0.6\textwidth]{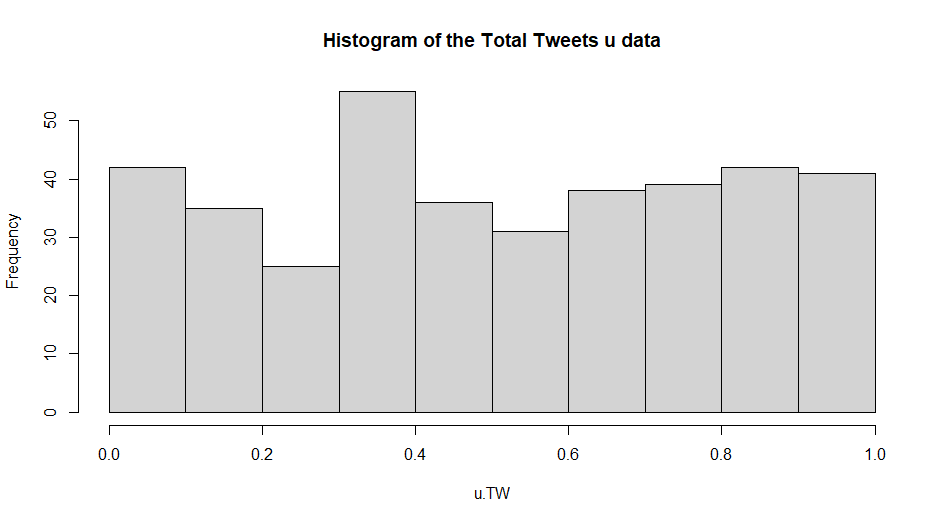}
    \caption{Plot illustrating the fit of the residuals for the \texttt{Tweets} marginal. Top plot: QQ-plot comparing the Gaussian theoretical quantiles with sample quantiles. Middle plot: observed time series (black line) and in-sample predictions obtained form the fitted SEP4 model (red line). Bottom plot: Histogram of the resulting \textit{u-data}.}
    \label{fig:qqplot_insample}
\end{figure}

As an example, Figure \ref{fig:qqplot_insample} shows the fit of the residuals for the \texttt{Tweets} marginal. The top panel displays the QQ-plot comparing the Gaussian theoretical quantiles with the sample quantiles, the middle panel illustrates the observations (black line) and in-sample predictions obtained from the fitted SEP4 model (red line), while the bottom panel shows the histogram of the resulting \textit{u-data}. The plots clearly show an excellent fit of the SEP4 model to the marginal, as demonstrated by the points in the QQ-plot aligning well to the main diagonal, the in-sample predictions overlapping the observed data and the shape of the \textit{u-data} histogram being close to a uniform pattern.

\subsection*{Vine Estimation}\label{sec:vine}

Once the marginals were estimated, we derived the corresponding \textit{u-data} from the residuals. %, as illustrated in Section \ref{sub:marginals}. 
Then, we carried out fitting and model selection for the vine copula. %using the R package \texttt{VineCopula} \cite{vinecopula}.

\begin{figure}[ht]
    \centering
    \includegraphics[width=\linewidth]{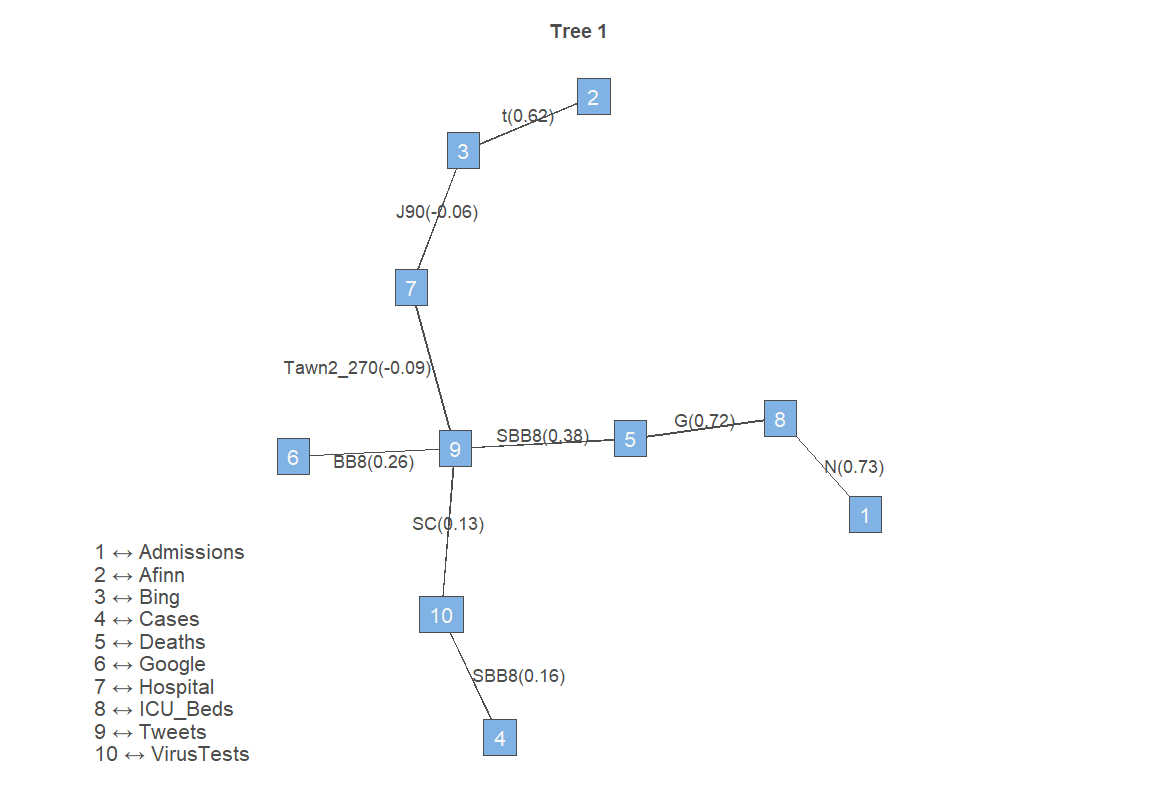}
    \caption{First tree of the vine copula for the Covid-19 data. The legend shows the names of the variables displayed on each node. The pair-copula families are shown on the edges and the Kendall's $\tau$s are given in brackets.}
    \label{fig:tree1}
\end{figure}

Figure \ref{fig:tree1} displays the first tree of the vine copula for the Covid-19 data, that we previously named model (i).
The nodes are denoted with blue squares, with the numbers corresponding to the margins reported on them.
On each edge, the plot shows the name of the selected pair copula family and the estimated copula parameter expressed as Kendall’s $\tau$. 
In order to estimate the vines, we adopted the Kendall’s $\tau$ criterion for tree selection, the AIC for the copula families selection and the MLE method for estimating the pair copula parameters. 
As it is clear from Figure \ref{fig:tree1}, the total number of tweets plays a central role in the vine, linking official Covid-19 to social media variables.
The total number of tweets and Google searches are contiguously related. 
Likewise, the sentiment scores \texttt{Bing} and \texttt{Afinn} are directly associated.
\texttt{Tweets} is also directly connected to the number of deaths, the total number of tests and the number of hospital cases.
The symmetric Gaussian copula, which is often employed in traditional multivariate modelling, was only identified once as the best fitting copula, to link the number of patients on ventilation and the number of new admissions. 
This suggests that model (i), with different pair copula families, better fits the data compared to model (ii), which assumes all Gaussian pair copula families.
On the contrary, the selected copula families include the Student’s t copula, which is able to model strong tail dependence, the Tawn copula and Archimedean copulas such as the Clayton, Gumbel and Joe, that are able to capture asymmetric dependence, and mixture copulas such as the BB8 (Joe-Frank), that can accommodate various dependence shapes. 
Most of the associations between the variables are positive. The strongest associations are between the official Covid-19 variables \texttt{ICU\_Beds} and \texttt{Admissions}, between \texttt{ICU\_Beds} and \texttt{Deaths} and between the \texttt{Bing} and \texttt{Afinn} sentiment scores. Also, \texttt{Deaths} and \texttt{Tweets} are mildly associated.
The results suggest that associations between official (a) and social media data (b) are present and that the independence model (iii), which assumes no relationship between different sources of information is not appropriate for the data.

\subsection*{Out-of-sample prediction results}

In this Section we constructed out-of-sample predictions using the proposed vine methodology (model (i)), which integrates official and social media Covid-19 variables.
We then compared the predictions obtained with our methodology with those yielded using two traditional approaches. 
The former is based on vines built exclusively using Gaussian pair copulas (model (ii)), which are the most common in applications, but are restricted to dependence symmetry and absence of tail dependence. The latter approach assumes independence among the ten time series under consideration (model (iii)) and therefore calculates predictions ignoring any association between official (a) and online information (b).
Since the vine approach allows all the variables to hold the same role, we calculated predictions for each variable in turn based on the remaining variables.

Out-of-sample predictions based on the proposed model were constructed %as illustrated in Section \ref{sec:out-of-sample}, 
considering the vine copula estimated 
%as explained in Section \ref{sec:vine} 
until the \nth{1} April 2021 and using it to predict the period between the \nth{2} April 2021 and the \nth{9} May 2021.  

Tables \ref{tab:mse} and \ref{tab:mis} list the MSE and MIS values calculated for each variable. 
The second columns show the vine copula (model (i)) results, the third columns show the results assuming all Gaussian pair-copulas (model (ii)) and the fourth columns show the results assuming independence among variables (model (iii)).
The MSEs and MISs of the best performing approaches for each variable are highlighted in boldface. 
Tables \ref{tab:mse} and \ref{tab:mis} show a similar model performance across the ten variables. 
According to both the MSE and MIS indicators, the vine copula approach (i) outperforms the other two approaches (ii) and (iii) for predicting the official (a) variables  \texttt{Deaths}, \texttt{Hospital} and \texttt{VirusTest}. 
The Gaussian vine approach (ii) also performs well with several variables, while the independent approach (iii) seems to exceed the other two approaches only with the variables \texttt{Admissions} and \texttt{Afinn}.

The 6 official (a) Covid-19 variables considered in this paper, \texttt{Admissions}, \texttt{Cases}, \texttt{Deaths}, \texttt{Hospital}, \texttt{ICU\_Beds} and \texttt{VirusTests}, are generally better predicted by the vine method (i), as opposed to the Gaussian (ii) and independence (iii) methods.
This last approach assumes no dependence between any of the variables involved in the model. Hence, this approach indicates the absence of any association between the official (a) and the social media (b) variables, implying the lack of contribution of online-generated information in predicting the official Covid-19 variables. 
On the contrary, the vine approach assumes the presence of a dependence structure between the variables and, in particular, between the official (a) and social media (b) insights. 
In particular, the highly flexible vine model (i) allows us to model asymmetric and tail dependence; while the Gaussian vine model (ii) only allows for symmetric and no tail dependence.
Therefore, the better performance of the vine (i) and (ii) compared to the independence model (iii) demonstrates usefulness of social media information in forecasting official Covid-19 variables.

The prediction of the 4 online-generated information (b) considered in this paper (\texttt{Afinn}, \texttt{Bing}, \texttt{Google} and \texttt{Tweets}) also benefits from data integration. 
Indeed, most of the social media variables are more accurately forecasted by the vine model, particularly the Gaussian one (model (ii)). This indicates that the Gaussian approach, characterized by a symmetric dependence structure, is flexible enough to model the social media variables.

\section*{Concluding Remarks}\label{sec:Conclusions}

In this paper, we propose a new methodology aimed at obtaining more accurate forecasts compared to traditional approaches, for variables measuring the Covid-19 dynamics.
The proposed methodology is based on the integration of Covid-19 variables collected from official UK sources with online generated social media insights, relevant to the same geographical area. 

%Together with official Covid-19 information related to infection counts and the pressure on the national health service, we also gathered Google Trends searches and Twitter microblogging messages involving keywords related to the Covid-19 pandemic. From the tweets, we considered the volume as well as the sentiment scores, to investigate the feelings of people towards the pandemic. Our methodology is based on vine copulas, which are able to model the dependence structure between the marginals, and thus to take advantage of the association between official Covid-19 and social media variables. We tested our approach calculating out-of-sample predictions and comparing the vine copula method with two traditional approaches: the first based on a vine constructed with all Gaussian copulas, and the second based on independence between variables. 
%The results show that the vine copula method outperforms the other two approaches for predicting the number of deaths, hospital admissions and tests, demonstrating that our methodology is able to leverage social media information to obtain more accurate predictions of Covid-19 effects than the other two approaches. In some cases, the Gaussian vine copula method is selected, showing that the vine data integration approach is still achieving the best performance, although some variables are less affected by asymmetries and tail dependence. 
The results show that the vine copula method generally outperforms two other traditional approaches (more precisely, the Gaussian vine copula approach and the independence approach) for predicting the official pandemic variables number of deaths, hospital admissions and tests, demonstrating that the proposed methodology is able to leverage social media information to obtain accurate predictions of Covid-19 effects.

%The proposed methodology will support policy makers to understand, monitor and combat the pandemic, assisting key medical and governmental actors to make informed decisions and to efficiently and effectively plan and allocate necessary resources.

%Further investigations including additional social media information will be the object of future work. 
%Also, the proposed approach could be extended using a 7 day rolling average in the model to adjust for the delay due to UK government figures being under reported at the weekend.
%Another extension will involve Bayesian inference, which would allow us to incorporate other information, such as experts’ opinion, in the model. In addition, the use of more sophisticated machine learning approaches could be envisaged for deriving the sentiment variables to improve the proposed methodology.

%%%%%%%%%%%%%%%%%%%%%%%%%%%%%%

\section*{Data Availability}

The datasets analysed during the current study are available in the GOV.UK repository, \url{https://coronavirus.data.gov.uk}, in the Johns Hopkins University repository, \url{https://coronavirus.jhu.edu/region/united-kingdom}, via the Twitter API, \url{https://developer.twitter.com}, and via the Google Trends API, \url{https://trends.google.com}.

%%%%%%%%%%%%%%%%%%%%%%%%%%%%%%%

\bibliography{sample}

%\noindent LaTeX formats citations and references automatically using the bibliography records in your .bib file, which you can edit via the project menu. Use the cite command for an inline citation, e.g.  \cite{Hao:gidmaps:2014}.

%For data citations of datasets uploaded to e.g. \emph{figshare}, please use the \verb|howpublished| option in the bib entry to specify the platform and the link, as in the \verb|Hao:gidmaps:2014| example in the sample bibliography file.

\section*{Acknowledgements}

This work was supported by EPSRC project \textit{Dependence Modelling with Vine Copulas for the Integration of Unstructured and Structured Data}, grant number EP/W021986/1, and by the ERDF project \textit{Environmental Futures \& Big Data Impact Lab}, funded by the ESIF, grant number 16R16P01302.

\section*{Author contributions statement}

L.A. collected the data, developed the code and implemented the analysis, L.D.V. supervised the project, wrote and edited the paper.
Both authors reviewed the manuscript. 

\section*{Additional information}

\textbf{Competing interests}:
The authors declare no competing interests.

%Figures and tables can be referenced in LaTeX using the ref command, e.g. Figure \ref{fig:stream} and Table \ref{tab:example}.

\begin{table}[ht]
\centering
\caption{MSEs calculated for each variable. The figures show the vine copula results (second column), the results assuming all Gaussian pair-copulas (third column), and assuming independence among variables (fourth column). The MSEs of the best performing approaches for each variable are in boldface.}
\begin{tabular}{|c|c|c|c|}
\hline
Marginal   & Vine Copula   & Gaussian  & Independent  \\
\hline
\texttt{Admissions} & 12510.57 & 12515.63 & \textbf{12506.15} \\
\texttt{Afinn} & 1.0226 & 1.0667 & \textbf{0.9766}\\
\texttt{Bing}   &  \textbf{0.2417}   &  0.2597  &  0.2452\\
\texttt{Cases} & 577363 & \textbf{577332.3} & 577401.1\\
\texttt{Deaths} & \textbf{886.7999} & 890.7506&887.5182\\
\texttt{Google}  &  385.3506 & \textbf{382.9412}   & 384.9532\\
\texttt{Hospital} & \textbf{1348411} & 1348573 & 1348496\\
\texttt{ICU\_Beds} & 49150.17 & \textbf{49142.98} & 49144.45\\
\texttt{Tweets}  & 17593.4  & \textbf{17583.2}   & 17585.62  \\
\texttt{VirusTests} & \textbf{292876} & 292943.3 & 292902.5\\
\hline
\end{tabular}\label{tab:mse}%
\end{table}

\begin{table}[ht]
\centering
\caption{MISs calculated for each variable. The figures show the vine copula results (second column), the results assuming all Gaussian pair-copulas (third column), and assuming independence among variables (fourth column). The MISs of the best performing approaches for each variable are in boldface.}
\begin{tabular}{|c|c|c|c|}
\hline
Marginal   & Vine Copula   & Gaussian  & Independent \\
\hline
\texttt{Admissions} & 25.5155 & 25.5208 & \textbf{25.5101} \\
\texttt{Afinn} & 0.2510 & 0.2587 & \textbf{0.2461}\\
\texttt{Bing}   &    0.1187 & 0.1231 & \textbf{0.1148}\\
\texttt{Cases} & 173.1329 & \textbf{173.1258} & 173.1414\\
\texttt{Deaths} & \textbf{6.5837} & 6.5991 & 6.5867\\
\texttt{Google}  &   4.5348 & \textbf{4.1436} & 4.1482 \\
\texttt{Hospital} & \textbf{257.4757} & 257.4917 & 257.4841 \\
\texttt{ICU\_Beds} & 49.3001 &\textbf{49.2965} & 49.2972\\
\texttt{Tweets}  &  28.9344 & \textbf{28.9253} & 28.9278\\
\texttt{VirusTests} & \textbf{120.5653} & 120.5799 & 120.5709\\
\hline
\end{tabular}\label{tab:mis}%
\end{table}

\end{document}